\documentclass[12pt]{elsarticle}

\usepackage{amsmath}
\usepackage{slashed}
\usepackage{graphicx}

\usepackage{amssymb}

\begin{document}

\begin{frontmatter}



\title{
\hfill {\small ALBERTA-THY-07-17}\\
\hfill {\small TTP17-027}\\[20mm]
{ Three-loop quark form factor at high energy: the leading mass corrections}}


\author[a]{Tao Liu\,}
\author[a,b]{Alexander A. Penin\,}
\author[c]{and Nikolai Zerf\,}

\address[a]{Department of Physics, University of Alberta,\\
  Edmonton AB T6G 2J1, Canada}
\address[b]{Institut f{\"u}r Theoretische Teilchenphysik, 
  Karlsruher Institut f\"ur Technologie (KIT),\\
  76128 Karlsruhe, Germany}
  \address[c]{Institut f{\"u}r Theoretische Physik, Universit\"at Heidelberg,\\
D-69120 Heidelberg, Deutschland
}

\begin{abstract}
We compute the leading mass corrections  to the high-energy behavior of the 
massive quark  vector form factor  to three loops in QCD in the 
double-logarithmic approximation.
\end{abstract}


\end{frontmatter}




\vspace*{12mm}
\newpage
The  vector  form factor of a quark is a crucial  building block in the 
perturbative analysis of many processes in quantum chromodynamics. It is also 
the  simplest  scattering amplitude which can be used to study the infrared 
structure of perturbative QCD. The form factors of  a massless quark have  been 
evaluated  through the  three-loop approximation 
\cite{Baikov:2009bg,Gehrmann:2010ue} and even to four loops in the leading-color 
approximation \cite{Lee:2016ixa}. For a massive quark however  only the two-loop 
result  is available so far \cite{Bernreuther:2004ih,Bernreuther:2005gw}.  The 
complete calculation of the three-loop corrections is quite a challenging 
problem for the existing computational techniques.  Only recently  the 
leading-color contribution of the planar three-loop Feynman diagrams has  been 
found analytically  in terms of Goncharov polylogarithms retaining the full 
dependence on the  quark mass $m_q$ \cite{Henn:2016tyf}.  At the same time the 
full mass dependence is often excessive  for  practical applications and proper 
expansion of the result  in a given kinematical region  could be  sufficient 
(see {\it e.g.} 
\cite{Feucht:2004rp,Penin:2005kf,Penin:2005eh,Bonciani:2007eh,Bonciani:2008ep}). 
In particular, in the high-energy limit the corrections to the form factor can 
be expanded  in a small ratio $\rho=m_q^2/Q^2$, where $Q_\mu$ is the  large 
momentum transfer. The resulting series is  asymptotic with the coefficients 
dominated by the double-logarithmic contribution enhanced by the second power of 
the large logarithm $\ln\rho$ per each power of the strong coupling constant 
$\alpha_s$. In the leading order of the  small-mass  expansion  the origin and  
structure of the ``Sudakov''  double logarithms have been established  long time 
ago \cite{Sudakov:1954sw,Frenkel:1976bj}. The analysis has been subsequently 
generalized to subleading  logarithms 
\cite{Mueller:1979ih,Collins:1980ih,Sen:1981sd} and the leading-power result for 
the massive  quark form factor  is currently known through three loops up to the 
${\cal O}(\alpha_s^3)$ nonlogarithmic contribution, which is only available in 
the leading-color approximation (see \cite{Ahmed:2017gyt} and references 
therein).  By contrast,  the logarithmic structure of the power suppressed terms 
is not well understood and currently is under study in various contexts 
\cite{Penin:2014msa,Melnikov:2016emg,Penin:2016wiw}.  In particular, the  
leading power  corrections to the form factor in QED  have been recently 
evaluated in the double-logarithmic approximation to all orders in the coupling 
constant \cite{Penin:2014msa}. The result  determines the abelian part of the 
corrections to the quark form factor.  In the present  paper we  complete the 
analysis of the  three-loop contribution by evaluating its nonabelian part  and 
derive the ${\cal O}(\rho\ln^6\!\!\rho\,\alpha^3_s)$  correction to the form 
factor in  QCD.

\begin{figure}[t]
\begin{center}
\includegraphics[width=2.8cm]{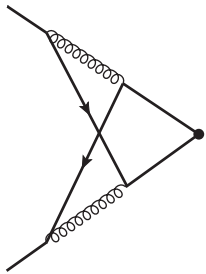}
\end{center}
\caption{\label{fig::2loop}  The two-loop diagram generating the 
${\cal O}(\rho)$ double-logarithmic contribution. The blob stands for the 
color singlet vector current.
}
\end{figure}

\begin{figure}[t]
\begin{center}
\begin{tabular}{ccc}
\includegraphics[width=2.5cm]{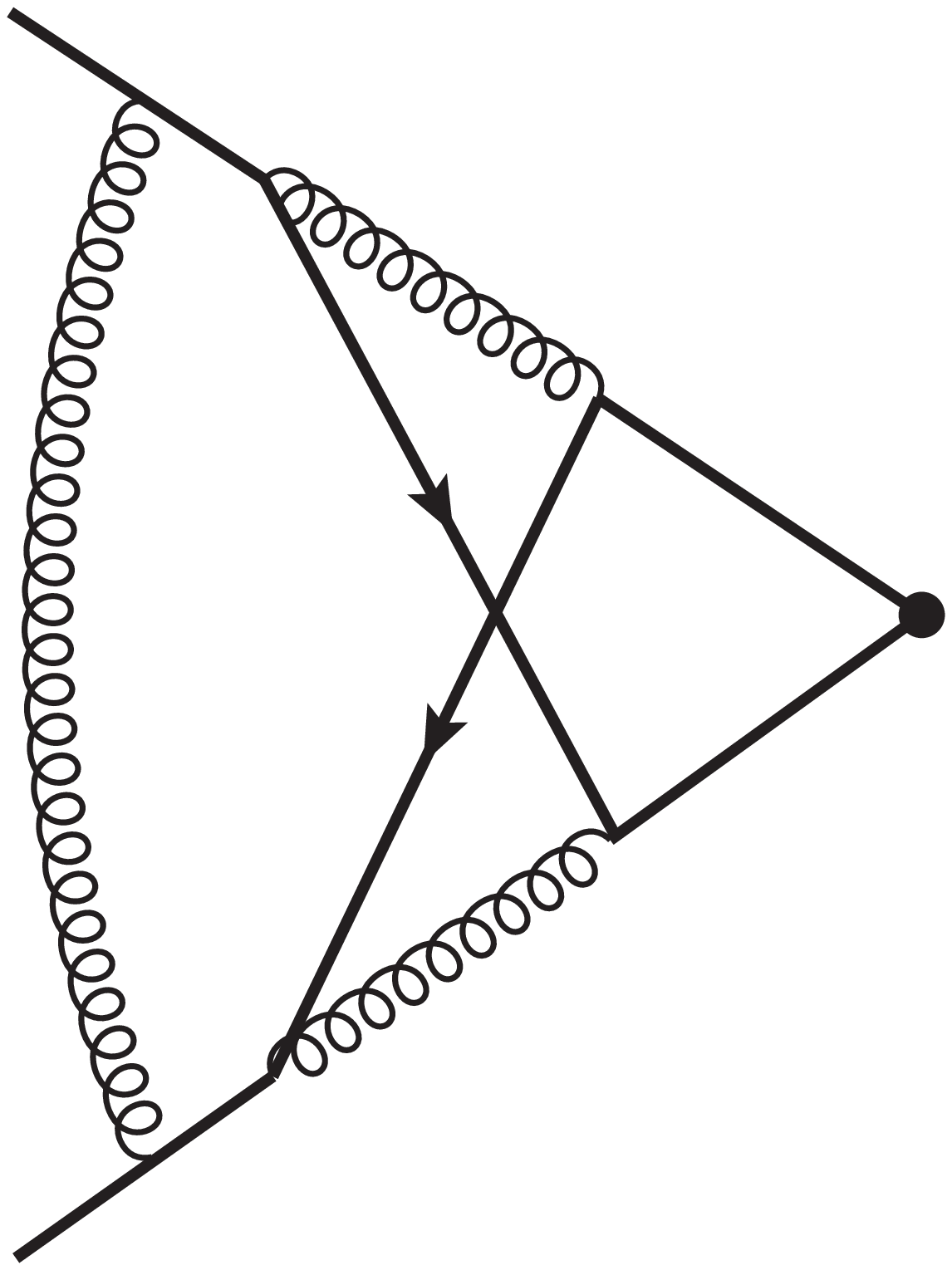}&
\hspace*{10mm}\includegraphics[width=2.5cm]{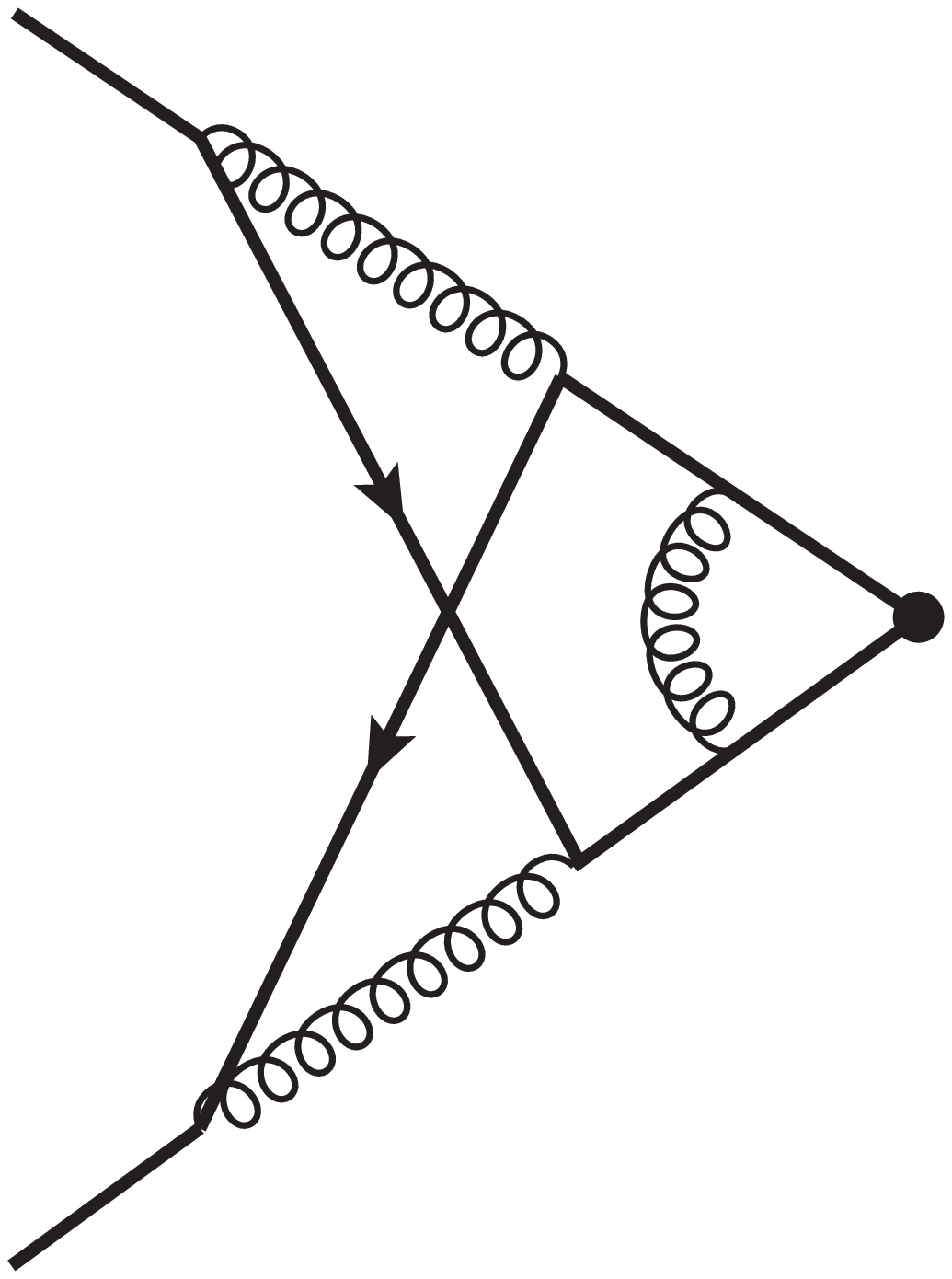}&
\hspace*{10mm}\includegraphics[width=2.5cm]{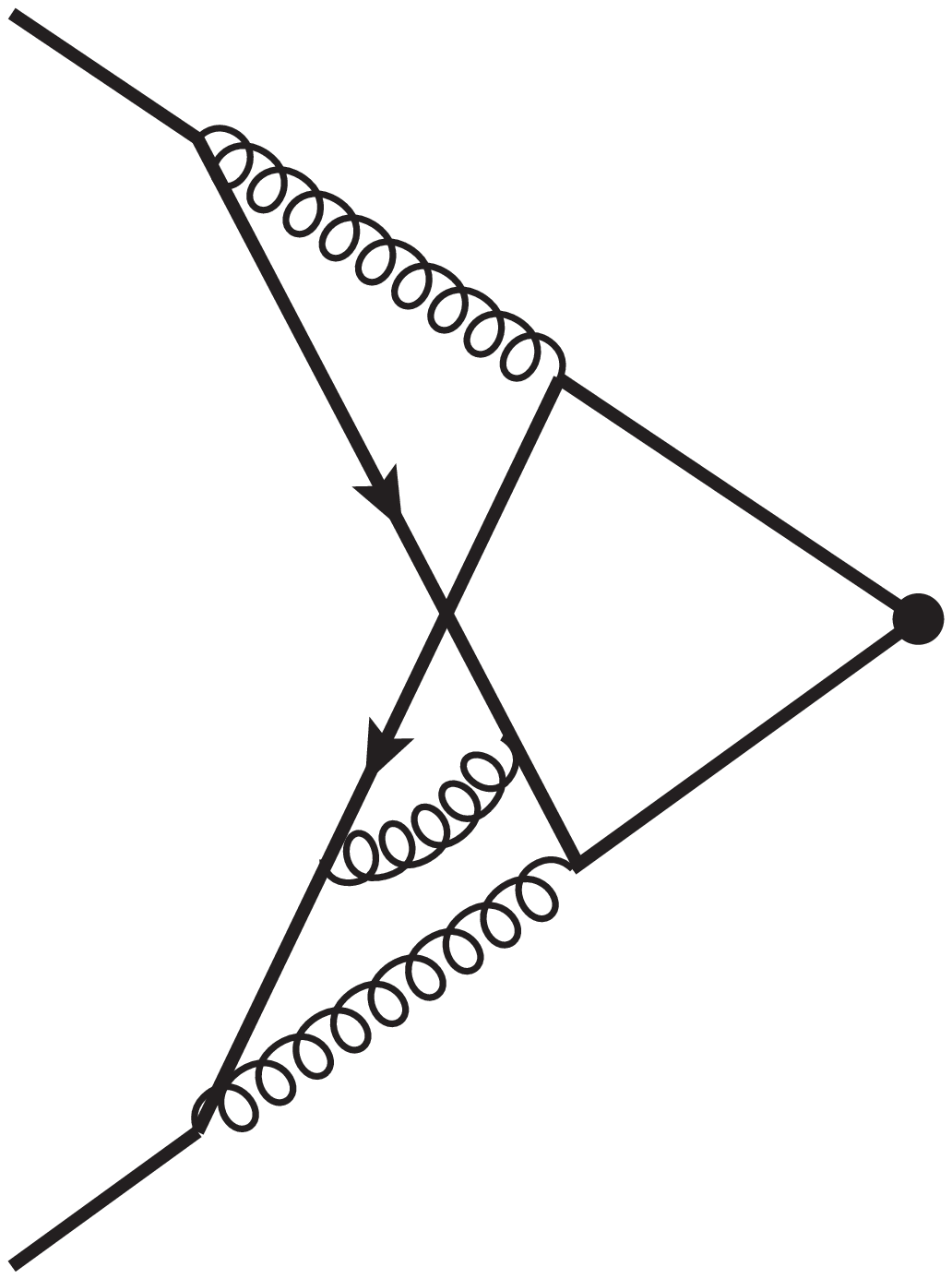}\\
(a)&\hspace*{10mm}(b)&\hspace*{10mm}(c)\\
\includegraphics[width=2.5cm]{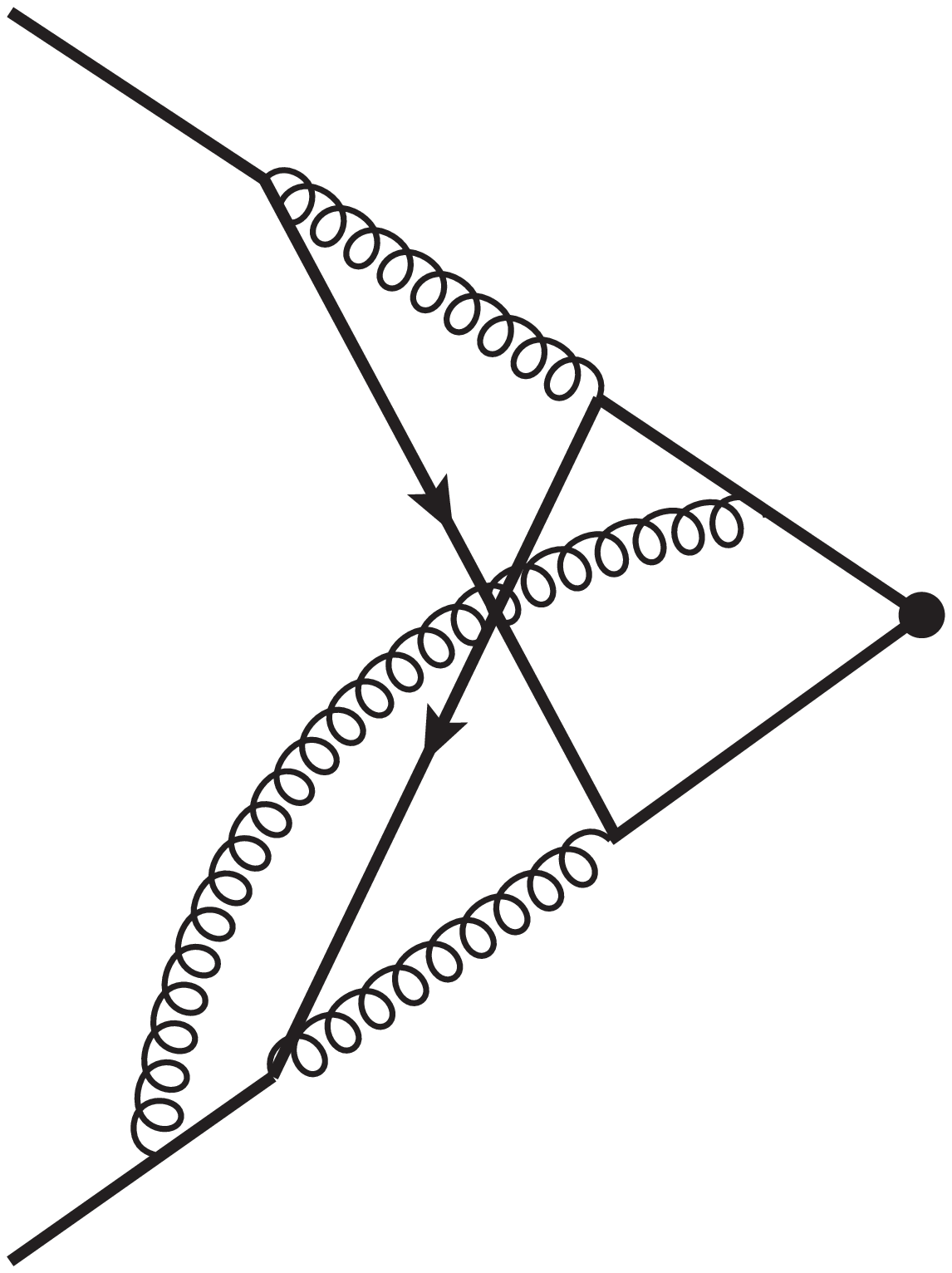}&
\hspace*{10mm}\includegraphics[width=2.5cm]{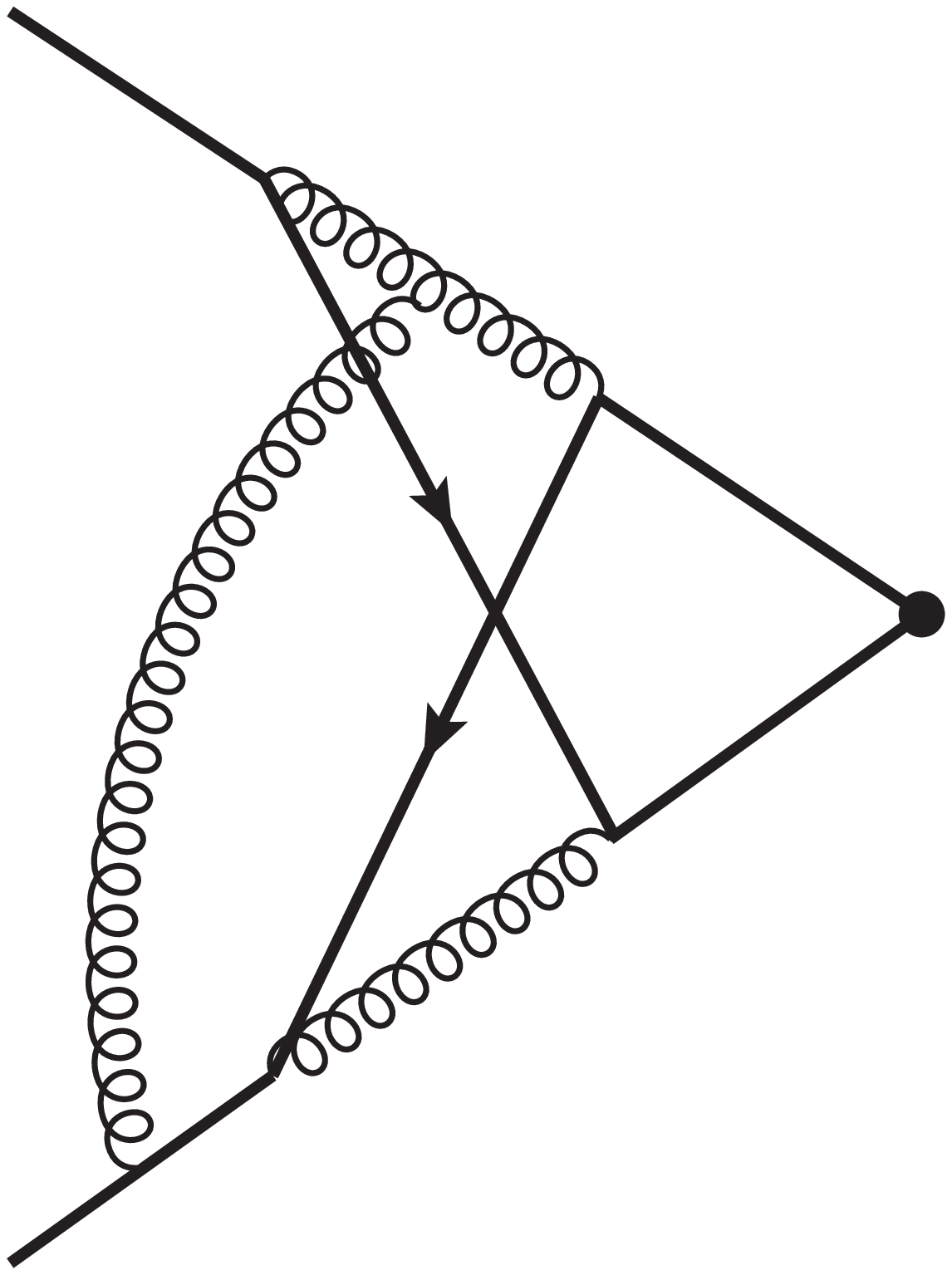}&
\hspace*{10mm}\includegraphics[width=2.5cm]{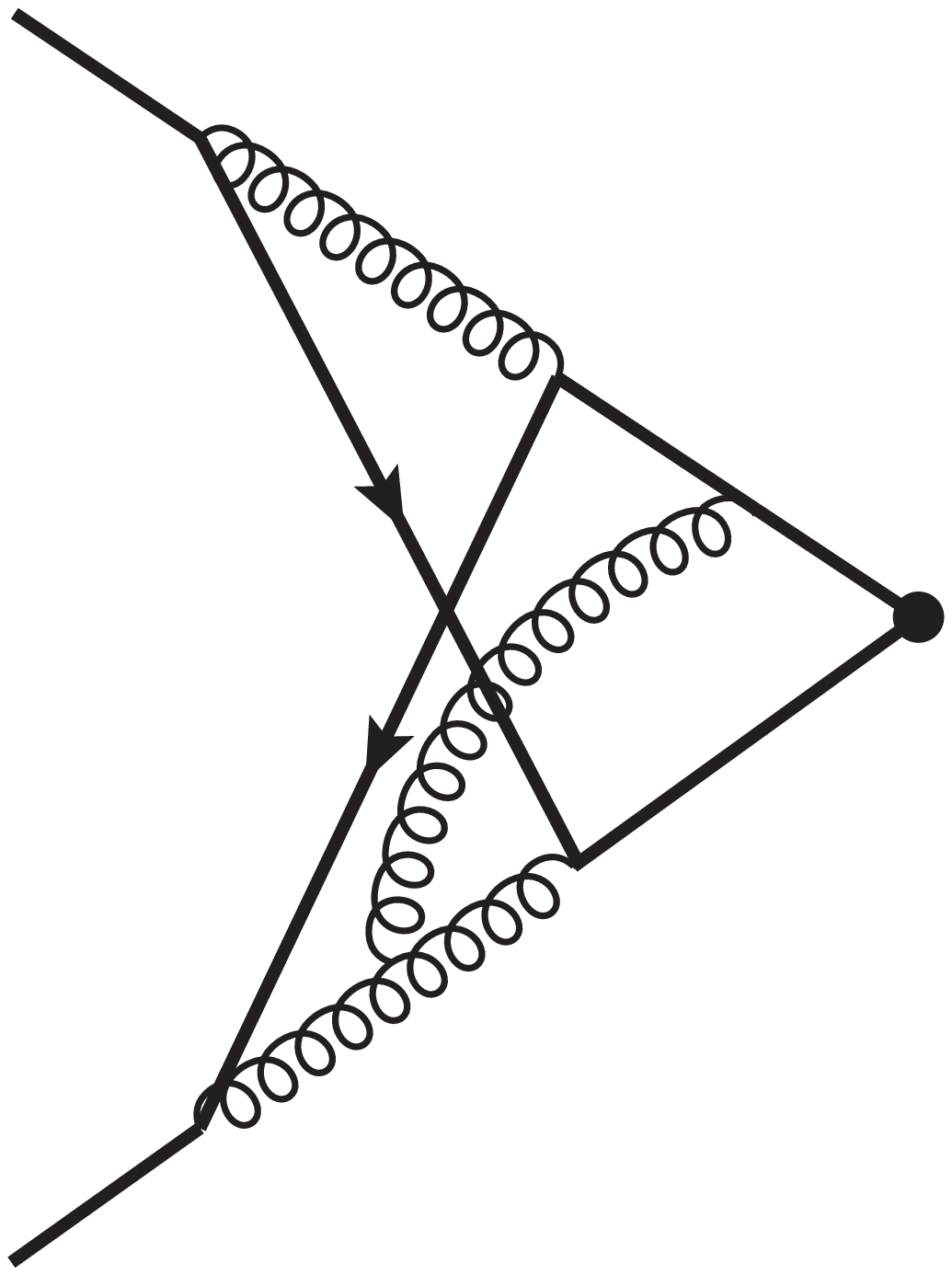}\\
(d)&\hspace*{10mm}(e)&\hspace*{10mm}(f)\\
\includegraphics[width=2.5cm]{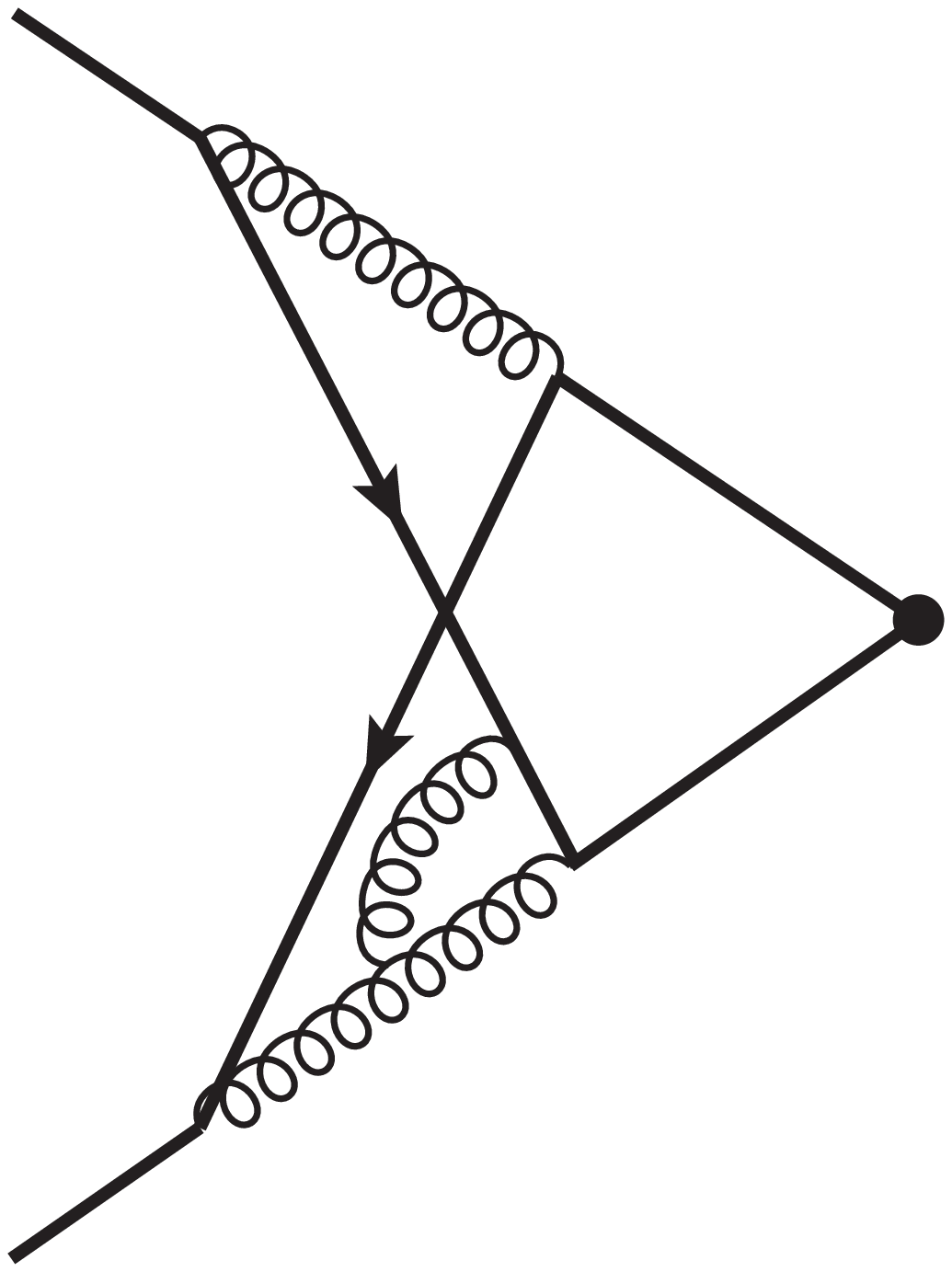}&
\hspace*{10mm}\includegraphics[width=2.5cm]{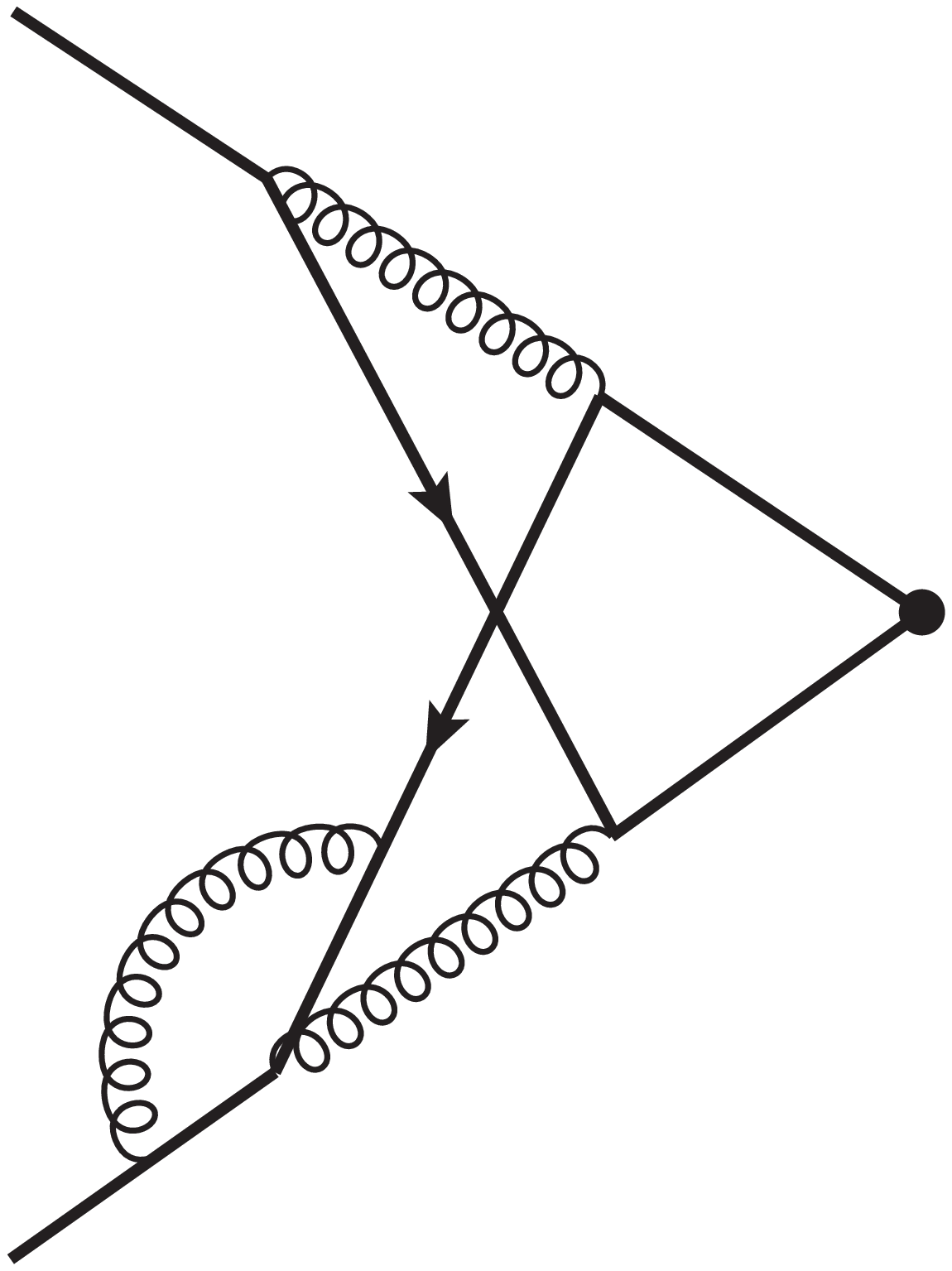}&
\hspace*{10mm}\includegraphics[width=2.5cm]{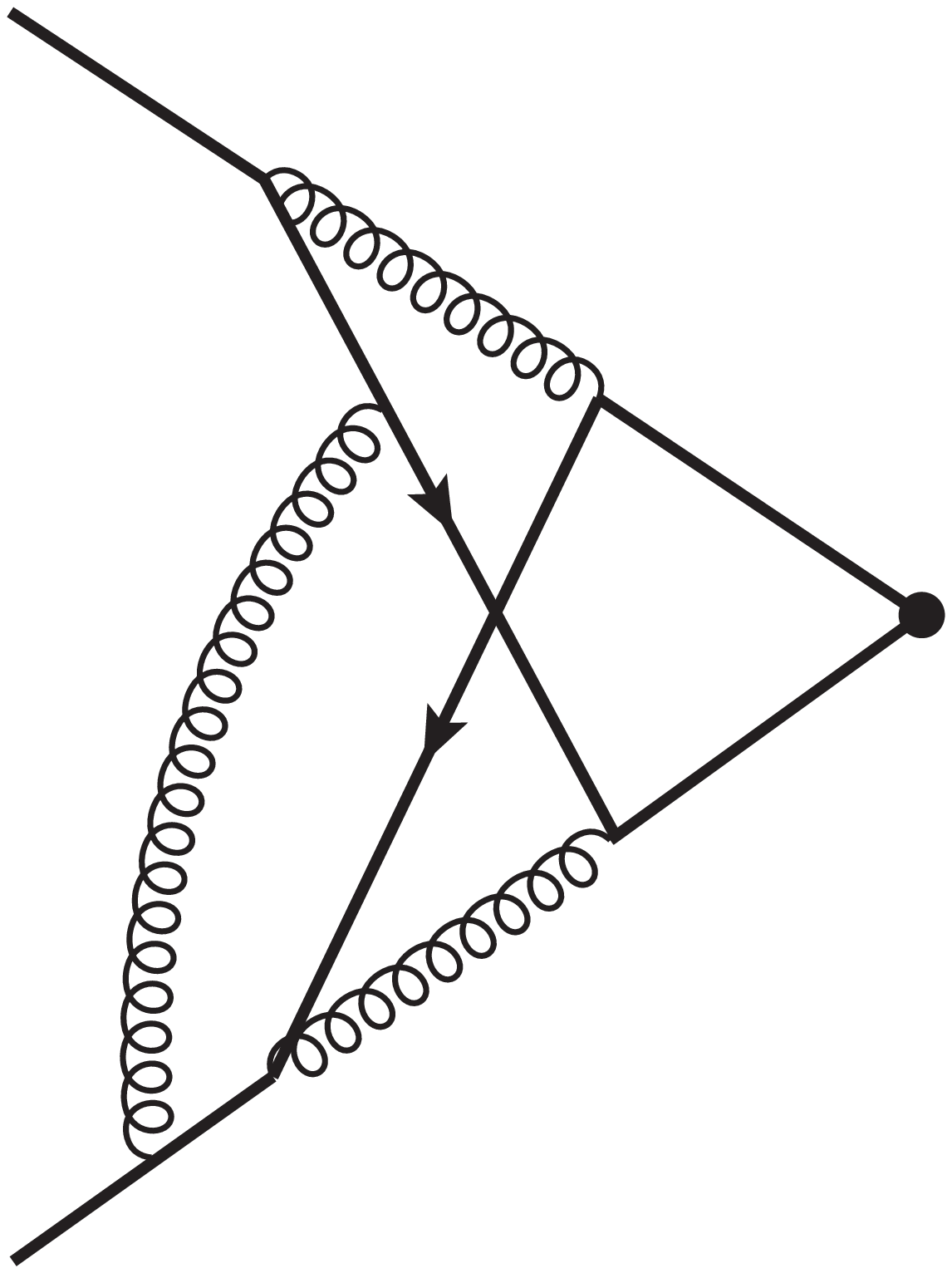}\\
(g)&\hspace*{10mm}(h)&\hspace*{10mm}(i)\\
\end{tabular}
\end{center}
\caption{\label{fig::3loop}  The three-loop diagrams contributing to 
the ${\cal O}(\rho)$ double-logarithmic corrections. Symmetric diagrams 
are not shown. The remaining diagrams either do not have the double-logarithmic 
integration region or have vanishing color factor.}
\end{figure}

The amplitude  ${\cal F}$  of the quark scattering in an external singlet vector 
field can be parametrized in the standard way  by the  Dirac and Pauli form 
factors
\begin{equation}
{\cal F}=\bar{q}(p_2)\left(\gamma_\mu F_1+{i\sigma_{\mu\nu}Q^\nu \over 2m_q}
F_2\right)q(p_1)\,.
\label{eq::Dirac}
\end{equation}
The Pauli form factor $F_2$ does not contribute in the approximation discussed 
in this paper and we focus on the high-energy behavior of the Dirac form 
factor $F_1$. We consider the on-shell quark $p_1^2=p_2^2=m_q^2$ and the large 
Euclidean momentum transfer $Q^2=-(p_2-p_1)^2$ corresponding to positive values 
of the parameter $\rho$. 
The asymptotic expansion of the Dirac form factor can be written as  follows
\begin{equation}
F_1=S_\varepsilon\sum_{n=0}^\infty \rho^n F^{(n)}_1\,,
\label{eq::F1}
\end{equation}
where $F^{(n)}_1$ are given by the power series in $\alpha_s$ with the
coefficients depending on $\rho$ only logarithmically.  The factor
\begin{equation}
S_\varepsilon=\exp{\left[-{\alpha_s\over 2\pi}\,
{\Gamma^{(1)}\over \varepsilon}\right]}
\label{eq::Sep}
\end{equation}
accounts for the singular dependence on the parameter of the dimensional 
regularization $d=4-2\varepsilon$ used  to treat  the infrared divergences of 
the amplitude. Here  $\Gamma^{(1)}$  is the one-loop cusp anomalous dimension. 
In the high-energy limit $\rho\to 0$ it reads \cite{Korchemsky:1987wg}
\begin{equation}
\Gamma^{(1)}=C_F\ln\rho\left(1+{\cal O}(\rho^2)\right),
\label{eq::Gam1}
\end{equation}
where $C_F={N_c^2-1\over 2N_c}$, $N_c=3$. In the double-logarithmic 
approximation the leading term is given by the Sudakov exponent  
\cite{Sudakov:1954sw,Frenkel:1976bj}
\begin{equation}
F^{(0)}_1=e^{-C_Fx}\,,
\end{equation}
where 
\begin{equation}
 x={\alpha_s\over 4\pi}\ln^2\rho
\end{equation}
is the double-logarithmic variable. The goal of this  paper is to compute the  
leading power  correction coefficient $F^{(1)}_1$ to ${\cal O}(x^3)$. The origin 
of the  ${\cal O}(\rho)$  double-logarithmic corrections is quite peculiar.  
They are induced by  the emission of soft virtual fermions rather than gauge 
bosons responsible for the  Sudakov logarithms 
\cite{Penin:2014msa,Penin:2016wiw}. The mass suppression factor in this case 
comes from  the helicity flip term in  the  soft fermion propagator, which 
effectively becomes scalar and is sufficiently singular at small momentum  to 
develop the double-logarithmic contribution.  In the case of the form factor the 
${\cal O}(\rho)$  double-logarithmic contribution is associated with the soft 
scalar quark pair exchange and appears first in the two-loop nonplanar vertex 
diagram,  Fig.~\ref{fig::2loop} \cite{Penin:2014msa}. The higher-order 
double-logarithmic corrections are obtained by dressing this diagram with extra 
soft gluons. The relevant three-loop diagrams are given in 
Fig.~\ref{fig::3loop}.

Let us briefly describe how the  diagrams are evaluated in the 
double-logarithmic approximation 
\cite{Penin:2014msa,Melnikov:2016emg,Penin:2016wiw}. Since two soft quark 
propagators provide the explicit mass suppression factor, the double logarithmic 
asymptotic of the integral over the virtual momenta can be obtained by the  
technique  originally applied to the analysis of the leading-power term  
\cite{Sudakov:1954sw}.  To introduce  the main idea  of the method we consider  
the evaluation of the two-loop diagram, Fig.~\ref{fig::2loop}. The 
double-logarithmic contribution originates from the  momentum configuration  
when the large external momenta flow through the edges of the diagram. In the 
infrared region all the propagators with the external momenta are eikonal and 
the edges of the diagram effectively turn into the light-cone Wilson lines.  At 
the same time the  momenta $l_i$ of the  exchanged quark pair are soft and the 
corresponding propagators in the infrared region become scalar. The effective 
Feynman rules for this momentum region, which retain  the leading infrared 
behavior of the full theory, are given in \cite{Penin:2016wiw}. To separate the 
double-logarithmic contribution the Sudakov  parametrization 
$l_i=u_ip_1+v_ip_2+{l_i}_\perp$ is used for  each virtual soft quark  momentum. 
The integration over the transverse components ${l_i}_\perp$ is performed by 
taking the residues of the soft propagators. In general the resulting expression 
has double-logarithmic scaling when $u_i,\,v_i\ll 1$ and the Sudakov parameters  
are ordered along the Wilson lines. For the nonplanar diagram under 
consideration this condition  reads $v_2\ll v_1 \ll 1$,  $u_1\ll u_2\ll 1$.  An 
additional constraint  $\rho\ll u_iv_i$  ensures that the soft quark propagators 
can go on-shell. This condition also suggests that $\rho\ll u_i,\,v_i$, which 
sets the infrared cutoff on the integral over the Sudakov parameters. Thus  the 
quark mass regulates both collinear and soft divergences and the result for the 
diagram is infrared finite.  In this way the two-loop contribution can be 
reduced to the following expression \cite{Penin:2014msa}\footnote{ The detailed 
derivation can be found in Ref.~\cite{Penin:2016wiw} in the context of two-loop 
analysis of Bhabha scattering. The relevant contribution is proportional to the 
integral $I_1$ in the Appendix A.} 
\begin{eqnarray}
F_1^{(1,2l)}&=&2\left(C_A-2C_F\right)x^2
\int K(\eta_1,\eta_2,\xi_1,\xi_2)
 {\rm d}\eta_1{\rm d}\eta_2{\rm d}\xi_1{\rm d}\xi_2\,,
\label{eq::F12loop}
\end{eqnarray}

\noindent 
where $C_A=N_c$, $\eta_i=\ln v_i/\ln\rho$ and  $\xi_i=\ln u_i/\ln\rho$ are the 
normalized logarithmic integration variables, the integration goes over the 
four-dimensional cube $0<\eta_i,~\xi_i<1$, and the kernel
\begin{eqnarray}
 K(\eta_1,\eta_2,\xi_1,\xi_2)&=&
 \theta(1-\eta_1-\xi_1)\theta(1-\eta_2-\xi_2)
 \theta(\eta_2-\eta_1) \theta(\xi_1-\xi_2)
 \label{eq::kernel2loop}
\end{eqnarray}
selects the kinematically allowed region of double-logarithmic integration 
discussed above. After integrating Eq.~(\ref{eq::F12loop}) one gets
\begin{equation}
 F^{(1,2l)}_1={C_F\left(C_A-2C_F\right)\over 6}x^2\,,
\label{eq::2loopres}
\end{equation}
in agreement with \cite{Bernreuther:2004ih}. 
The three-loop correction can be represented as a sum over
the contribution of the diagrams in  Fig.~\ref{fig::3loop}
\begin{equation}
 F^{(1,3l)}_1={{C_F}\left(C_A-2C_F\right)\over 2}
 \sum_\lambda c_\lambda d_\lambda \,x^3\,,
\label{eq::F13loop}
 \end{equation}
where the diagrams $(d)$-$(i)$ with a symmetric counterpart should be counted 
twice. Here $c_\lambda$ stands for a reduced color factor and the three-loop
integrals are converted into the  following form  
\begin{equation}
d_\lambda=4\int w_\lambda(\eta,\xi)K(\eta_1,\eta_2,\xi_1,\xi_2)
 {\rm d}\eta_1{\rm d}\eta_2{\rm d}\xi_1{\rm d}\xi_2\,,
\label{eq::dlam}
 \end{equation}
where $w_\lambda$ is the weight function resulting from the double-logarithmic 
integration over the  soft gluon momentum.  The results   for $w_\lambda$, 
$d_\lambda$, and  $c_\lambda$ are listed in Table~\ref{tab::1}.  Examples of the 
calculation of the functions  $w_\lambda$  are given in the~\ref{sec::A}. Note 
that the diagram Fig. 2\,(a) has an infrared divergent contribution which 
reproduces the factorized singular structure of Eq.~(\ref{eq::F1}) and is not 
included into Eq.~(\ref{eq::F13loop}).

\begin{table}[t]
\begin{center}
    \begin{tabular}{|c|c|c|c|}
     \hline
       $\lambda$ &  $w_\lambda$  & $d_\lambda$ & $c_\lambda$ \\
    \hline
      ~a~  & $-((\eta_2 + 2)\eta_2+(\xi_1-2\eta_2+2)\xi_1-1) $ & $-\frac{17}{45}$ &   $-C_F$ \\
      b & $2\xi_2\eta_1$                                  &$\frac{1}{45}$ &   $-C_F$ \\
      c & $2(\xi_1-\xi_2)(\eta_2-\eta_1)$              &$\frac{1}{15}$ &   $C_A-C_F$ \\    
      d & $-\eta_1(\eta_1-2\xi_1+2) $                     &$-\frac{1}{10}$ &  $C_A-C_F$ \\
      e & $(\eta_2-\eta_1)(2 - 2\xi_1 + \eta_1 + \eta_2)$ &$\frac{8}{45}$ &   $-\frac{C_A}{2}$ \\
      f & $2\eta_1(\xi_1 - \xi_2)$                     &$\frac{1}{30}$ &   $-\frac{C_A}{2}$ \\     
      g & $2\eta_2(\xi_1-\xi_2)$                       &$\frac{1}{10}$ &   $-\frac{C_A}{2}$ \\
      h & $\eta_1(\eta_1-2\xi_1+2) $                     &$\frac{1}{10}$ &   $\frac{C_A}{2}-C_F$  \\
      i & $\eta_2(\eta_2-2\xi_1+2)$                      &$\frac{5}{18}$ &   $\frac{C_A}{2}-C_F$  \\  
    \hline
    \end{tabular}  
\end{center}    
    \caption{\label{tab::1}
      The weights $w_\lambda$, integrals $d_\lambda$, and color factors 
      $c_\lambda$ of the diagrams in Fig.~\ref{fig::3loop}. To obtain $w_a$ the 
      singilar part of the infrared divergent diagram (a) is subtracted as discussed 
      in the~\ref{sec::A}.}
\end{table}

Collecting the contributions of the individual diagrams we get
\begin{eqnarray}
F^{(1,3l)}_1&=&{8C_F^3-2C_AC_F^2-C_A^2C_F\over 30}x^3
\label{eq::3lresult}
\end{eqnarray}
and 
\begin{eqnarray}
F^{(1)}_1&=&{C_F\left(C_A-2C_F\right)\over 6}x^2\left[1-
{C_A+4C_F\over 5}x \right.
+{\cal O}(x^2)\bigg].
\label{eq::final}
\end{eqnarray}
Thus, we have evaluated the dominant power  corrections to the tree-loop massive 
quark vector form factor at high energy. Only the nonplanar diagrams contribute 
to Eq.~(\ref{eq::final}) and the result has the subleading color  factor 
$C_A-2C_F$ which scales as $1/N_c$ in the large $N_c$ limit. This agrees with  
the leading-color analysis of  Ref.~\cite{Henn:2016tyf}, where such term is 
absent and the  ${\cal O}(\rho\,\alpha^3_s)$ contribution has at most the fifth 
power of the large logarithm. Our result can be used as a cross check for the 
future exact calculation of the three-loop corrections.  It can be used also to 
identify and extend the domain where the high energy approximation 
\cite{Ahmed:2017gyt} is applicable. An interesting and important problem is to 
extend Eq.~(\ref{eq::final}) to all orders in $x$. So far all-order resummation 
of the non-Sudakov double logarithms has been performed  only in abelian gauge 
theories \cite{Penin:2014msa,Melnikov:2016emg,Gorshkov:1966ht}. Generalization 
of the analysis to the nonabelian case can be crucial in particular for the 
analysis of the light quark effects in Higgs boson production 
\cite{Melnikov:2016emg}, and our result can be considered as the first step 
towards this goal.

\section*{Acknowledgments}
We would like to thank  M. Steinhauser for careful reading the manuscript and 
useful comments. The work of A.P. is supported in part by NSERC and  Perimeter 
Institute for Theoretical Physics. The work of T.L. is supported by NSERC.

\vspace{10mm}

\appendix
\section{Evaluation of the soft gluon momentum integrals}
\label{sec::A}
Besides the integration over two soft quark momenta, the three-loop diagrams 
include an extra integration over the  soft gluon momentum. In general, this 
integration  can  be performed in the  double-logarithmic approximation within 
the Sudakov method  outlined above. However, the  analysis of the diagrams with 
the soft gluon emission from the on-shell external or soft quark lines is more 
subtle  due to soft divergences which are not regulated by the quark mass as in 
the two-loop case. We describe how this problem is treated  for the two typical 
cases of the diagrams (a) and (h) in Fig. 2. 

Fig. 2\,(a) is the only diagram with the infrared divergence in the final  
result.  The integration over the soft gluon momentum $l_3$ in this diagram is 
double-logarithmic when  one can neglect it in the eikonal propagators with the  
the soft  quark momenta $l_{1,2}$. This defines  the conditions $l_3p_1\ll 
l_2p_1$,  $l_3p_2\ll l_1p_2$  corresponding to the ordering of the Sudakov 
parameters $v_3\ll v_2$, $u_3\ll u_1$.  Thus $l_3$ should be retained only in 
the propagators without the soft quark momenta and the integral over the soft 
gluon momentum is reduced to 
\begin{equation}
{2iQ^2\over \pi^2 }\int{{d^4l_3}\over l_3^2 ((p_1+l_3)^2-m_q^2) 
((p_2 +l_3)^2-m_q^2)}\,,
\label{eq::intl3}
\end{equation}
with the above restriction on $l_3$ and the prefactor introduced for 
convenience. In the double-logarithmic approximation the propagators in this 
expression take the following form
\begin{eqnarray}
&& {1\over l_3^2}  \approx
- i \pi \delta(Q^2u_3v_3 + {l_3}_\perp^2)\,,
\nonumber\\
&&{1\over (p_1+l_3)^2-m_q^2}\approx \frac{1}{Q^2(v_3+2\rho u_3)}\,,
\nonumber\\
&&{1\over (p_2 +l_3)^2-m_q^2} \approx \frac{1}{Q^2(u_3+2\rho v_3)}\,.
\label{eq::prop}
\end{eqnarray}
After integrating  Eq.~(\ref{eq::intl3}) over ${l_3}_\perp$  with the 
double-logarithmic accuracy we get  
\begin{equation}
2\int_{\rho u_3}^{v_2}{{\rm d}v_3\over v_3}
\int_{\rho v_3}^{u_1}{{\rm d}u_3\over u_3}\,.
\label{eq::intluv}
\end{equation}
Eq.~(\ref{eq::intluv})  has soft  divergence  when $v_3$ and $u_3$ 
simultaneously become small.  This divergence can be removed by subtracting 
the factorized expression
\begin{equation}
2\int_{\rho u_3}^{1}{{\rm d}v_3\over v_3}
\int_{\rho v_3}^{1}{{\rm d}u_3\over u_3}\,.
\label{eq::intlfac}
\end{equation}
The subtraction term does not depend on the soft quark momenta.  It  is 
equivalent to the double-logarithmic approximation of the one-loop correction to 
the form factor and gives the following contribution to Eq.~(\ref{eq::F1})
\begin{equation}
 -\left({\alpha_s\over 2\pi}\,
{\Gamma^{(1)}\over \varepsilon} +C_Fx\right)\rho F_1^{(1,2l)}\,.
\label{eq::intsing}
\end{equation}
The first term of Eq.~(\ref{eq::intsing}) reproduces the singular ${\cal 
O}(\rho\alpha_s^3)$  part of  Eq.~(\ref{eq::F1}) while  the second term should 
be included in Eq.~(\ref{eq::F13loop}). The  subtracted expression reads
\begin{eqnarray}
&-&
2\left(\int_{v_2}^{1}{{\rm d}v_3\over v_3}\int_{\rho v_3}^{u_1}{{\rm d}u_3\over u_3}
+\int_{\rho u_3}^{v_2}{{\rm d}v_3\over v_3}\int_{u_1}^{1}{{\rm d}u_3\over u_3}
+\int_{v_2}^{1}{{\rm d}v_3\over v_3}
\int_{u_1}^{1}{{\rm d}u_3\over u_3}\right)
\nonumber \\
&& \nonumber \\
&=& -\left(
\ln v_2\left(\ln v_2+2\ln\rho\right)
+\ln u_1\left(\ln u_1-2\ln v_2 +2\ln\rho\right) \right)
\,.
\label{eq::intsub}
\end{eqnarray}
After converting  to the logarithmic variables the above equation together with 
the nonsingular term of Eq.~(\ref{eq::intsing}) gives the expression for 
$w_a$ in Table.~\ref{tab::1}. 

A characteristic feature of the diagram  Fig.~2\,(h) is that the soft gluon is 
emitted by a soft quark. In this case the Sudakov parametrization of its 
virtual momentum should be defined with respect to the corresponding soft quark 
momentum $l_3=u_3l_1+v_3p_2+{l_3}_\perp$. As in the two-loop contribution the 
integration  over the transverse component of  $l_2$ is performed by taking the 
residue of a soft quark propagator pole and there  exist two contributions 
corresponding to the on-shell propagators on either side of the soft gluon 
emission vertex.  When in Fig.~2\,(h) the soft quark propagator above the vertex 
is on the mass shell, the soft gluon momentum has to flow through the quark 
propagator below the vertex and the integral over $l_3$ coincides with the 
one-loop correction to the on-shell  form factor with the external momenta $p_1$ 
and  $l_2$. Using the same normalization as in Eq.~(\ref{eq::intl3}) it can be 
written as follows
\begin{eqnarray}
&&-{2i(p_2-l_1)^2\over \pi^2 }\int{{d^4l_3}\over 
l_3^2 ((p_2+l_3)^2-m_q^2)((l_1 +l_3)^2-m_q^2)}
\label{eq::intl31}
\end{eqnarray}
and in the standard way reduces to the integral over the Sudakov parameters
\begin{equation}
2\int_{\rho u_3/u_1}^{1}{{\rm d}v_3\over v_3}
\int_{\rho v_3/u_1}^{1}{{\rm d}u_3\over u_3}\,,
\label{eq::intluv1}
\end{equation}
where we used the relation $(p_2-l_1)^2\approx -Q^2u_1$. When in Fig.~2\,(h) the 
 soft quark propagator below  the vertex is on the mass shell, the soft gluon 
momentum has to flow through the quark propagator above  the vertex and instead 
of Eq.~(\ref{eq::intl31})  one gets  
\begin{eqnarray}
&&-{2i(p_2-l_1)^2\over \pi^2 }\int{{d^4l_3}\over 
l_3^2 ((p_2+l_3)^2-m_q^2)((l_1 - l_3)^2-m_q^2)}\,,
\label{eq::intl32}
\end{eqnarray}
with an additional condition  $p_1l_3\ll p_1l_1$ or   $v_3\ll v_1$ on the 
double-logarithmic integration region. This gives 
\begin{equation}
-2\int_{\rho u_3/u_1}^{v_1}{{\rm d}v_3\over v_3}
\int_{\rho v_3/u_1}^{1}{{\rm d}u_3\over u_3}\,.
\label{eq::intluv2}
\end{equation}
Both Eq.~(\ref{eq::intluv1}) and  Eq.~(\ref{eq::intluv2}) are infrared divergent. 
However, their  sum 
\begin{equation}
2\int_{v_1}^{1}{{\rm d}v_3\over v_3}
\int_{\rho v_3/u_1}^{1}{{\rm d}u_3\over u_3}
=\ln v_1\left(\ln v_1-2\ln u_1+2\ln\rho\right)
\label{eq::intluvsum}
\end{equation}
is finite and after converting  to the logarithmic variables coincides with the 
expression for  $w_h$ in Table.~\ref{tab::1}.

The evaluation of the rest of the diagrams  poses no new technical problem and 
can be performed in the same way.

\vspace{10mm}



\begin{thebibliography}{99}

\bibitem{Baikov:2009bg}
  P.~A.~Baikov, K.~G.~Chetyrkin, A.~V.~Smirnov, V.~A.~Smirnov and M.~Steinhauser,
  Phys.\ Rev.\ Lett.\  {\bf 102}, 212002 (2009).

\bibitem{Gehrmann:2010ue}
  T.~Gehrmann, E.~W.~N.~Glover, T.~Huber, N.~Ikizlerli and C.~Studerus,
  JHEP {\bf 1006}, 094 (2010).
 
 
\bibitem{Lee:2016ixa} 
  J.~Henn, A.~V.~Smirnov, V.~A.~Smirnov, M.~Steinhauser and R.~N.~Lee,
  JHEP {\bf 1703}, 139 (2017).
 

\bibitem{Bernreuther:2004ih}
  W.~Bernreuther, R.~Bonciani, T.~Gehrmann, R.~Heinesch, T.~Leineweber,
  P.~Mastrolia and E.~Remiddi,
  Nucl.\ Phys.\ B {\bf 706}, 245 (2005).


\bibitem{Bernreuther:2005gw}
  W.~Bernreuther, R.~Bonciani, T.~Gehrmann, R.~Heinesch, P.~Mastrolia and
  E.~Remiddi,
  Phys.\ Rev.\ D {\bf 72}, 096002 (2005).
  



\bibitem{Henn:2016tyf}
  J.~Henn, A.~V.~Smirnov, V.~A.~Smirnov and M.~Steinhauser,
  JHEP {\bf 1701}, 074 (2017).


  
\bibitem{Feucht:2004rp}
  B.~Feucht, J.~H.~Kuhn, A.~A.~Penin and V.~A.~Smirnov,
  Phys.\ Rev.\ Lett.\  {\bf 93}, 101802 (2004).

  
\bibitem{Penin:2005kf}
  A.~A.~Penin,
  Phys.\ Rev.\ Lett.\  {\bf 95}, 010408 (2005).


\bibitem{Penin:2005eh}
  A.~A.~Penin,
  Nucl.\ Phys.\ B {\bf 734}, 185 (2006).


\bibitem{Bonciani:2007eh}
  R.~Bonciani, A.~Ferroglia and A.~A.~Penin,
  Phys.\ Rev.\ Lett.\  {\bf 100}, 131601 (2008).

\bibitem{Bonciani:2008ep}
  R.~Bonciani, A.~Ferroglia and A.~A.~Penin,
  JHEP {\bf 0802}, 080 (2008).



\bibitem{Sudakov:1954sw}
  V.~V.~Sudakov,
  Sov.\ Phys.\ JETP {\bf 3}, 65 (1956)
  [Zh.\ Eksp.\ Teor.\ Fiz.\  {\bf 30}, 87 (1956)].


\bibitem{Frenkel:1976bj}
  J.~Frenkel and J.~C.~Taylor,
  Nucl.\ Phys.\ B {\bf 116}, 185 (1976).

  
\bibitem{Mueller:1979ih}
  A.~H.~Mueller,
  Phys.\ Rev.\ D {\bf 20}, 2037 (1979).

\bibitem{Collins:1980ih}
  J.~C.~Collins,
  Phys.\ Rev.\ D {\bf 22}, 1478 (1980).

\bibitem{Sen:1981sd}
  A.~Sen,
  Phys.\ Rev.\ D {\bf 24}, 3281 (1981).
 
   
\bibitem{Ahmed:2017gyt} 
  T.~Ahmed, J.~M.~Henn and M.~Steinhauser,
  arXiv:1704.07846 [hep-ph].
  
 
  
\bibitem{Penin:2014msa}
  A.~A.~Penin,
  Phys.\ Lett.\ B {\bf 745}, 69 (2015)
  Corrigendum: {\it ibid.} to be published, 
  [arXiv:1412.0671v4  [hep-ph]].
  

\bibitem{Melnikov:2016emg}
  K.~Melnikov and A.~Penin,
  JHEP {\bf 1605}, 172 (2016).

  
\bibitem{Penin:2016wiw}
  A.~A.~Penin and N.~Zerf,
  Phys.\ Lett.\ B {\bf 760}, 816 (2016).



\bibitem{Korchemsky:1987wg}
  G.~P.~Korchemsky and A.~V.~Radyushkin,
  Nucl.\ Phys.\ B {\bf 283}, 342 (1987).


\bibitem{Gorshkov:1966ht}
  V.~G.~Gorshkov, V.~N.~Gribov, L.~N.~Lipatov and G.~V.~Frolov,
  Sov.\ J.\ Nucl.\ Phys.\  {\bf 6}, 95 (1968)
  [Yad.\ Fiz.\  {\bf 6}, 129 (1967)].

\end{thebibliography}
\end{document}